# "Modeling Cosmic Expansion, and Possible Inflation, As a Thermodynamic Heat Engine."

By:   Dr. Christopher Pilot
      Physics Dept.
      Gonzaga University
      Spokane, WA 99258

      pilot@gonzaga.edu

**Abstract:**

If we assume a closed universe with slight positive curvature, cosmic expansion can modeled as a heat engine where we define the "system", collectively, as those regions of space within the observable universe, which will later evolve into voids/ empty space. We identify the "surroundings", collectively, as those pockets of space, which will eventually develop into matter-filled galaxies, clusters, super-clusters and filament walls. Using this model, we can find the energy needed for cosmic expansion using basic thermodynamic principles, and prove that cosmic expansion had as its origin, a *finite* initial energy density, pressure, volume, and temperature.  Inflation in the traditional sense, with the inflaton field, may also not be required.   We will argue that homogeneities and in-homogeneities in the WMAP temperature profile is attributable to quantum mechanical fluctuations about a fixed background temperature in the initial *isothermal* expansion phase.  Fluctuations in temperature can cause certain regions of space to lose heat.  Other regions will absorb that heat.  The voids are those regions which absorb the heat forcing, i.e., fueling expansion of the latter and creating slightly cooler temperatures in the former, where matter will later congregate.   Upon freeze-out, this could produce the observed WMAP signature with its associated CBR fluctuation in magnitude.  Finally, we estimate that the freeze-out temperature and the freeze-out time for WMAP in-homogeneities, occurred at roughly $3.02 * 10^{27}$ K and $2.54 * 10^{-35}$ s, respectively, after first initiation of volume expansion.  This is in line with current estimates for the end of the inflationary epoch.  The heat input in the inflationary phase is estimated to be $Q = 1.81 * 10^{94}$ J, and the void volume increases by a factor of only 5.65. The bubble voids in the observable universe increase, collectively, in size from about .046 $m^3$ to .262 $m^3$ within this inflationary period.

**Introduction:**

WMAP and Planck satellite data have confirmed that the universe as a whole is remarkably flat, within .5% of being perfectly flat. Early WMAP resolutions of the CMB, combined with other astronomical data, indicated a sum of density parameters equal to $1.02 \pm .02$ at the $1\sigma$ level [1]. This would indicate a slight positively curved space, compatible with strict flat space. More recent estimates [2-4] set the sum closer to $1.005^{+.016}_{-.017}$, which also seems to favor a slight positive curvature. That being said, it cannot be ruled out that the universe has no small curvature, either positive or negative. We assume an almost flat but closed universe, i.e., one with positive curvature, which will allow for eventual contraction and an ultimate so-called "big-bounce" scenario. Einstein already in 1917 understood and appreciated the unique advantage of a spatially closed cosmos in that it avoids the problem of trying to define a boundary [5,6] for the cosmos as a whole.

Cosmic inflation has been described [7] as the trillion, trillion-fold expansion of space within a trillionth-trillionth of a second, i.e., the entire universe expanded from the size of a proton to the size of a grapefruit within $10^{-32}$ s. Most theories of inflation do not distinguish between "system" and "surroundings", and they assume that the entire universe, or the entire observable part of the universe, inflates. While the inflation of space does not, per se, violate causality because there is no exchange of energy/information between material bodies, it is not clear unless one distinguishes between "system" and "surroundings" what is inflating and what is not. If inflation addresses the entire universe, or the entire observable universe, then it seems to us that causality is violated because the entire universe expands at a rate greater than the speed of light. If inflation addresses only the part, which expands, namely, the bubble voids, which is what will ultimately scale in our model, then we believe that causality will remain intact for the surroundings provided we are within the Hubble radius. In other words, there will be no restriction on how quickly the bubble voids inflate within our model. We also keep in mind that we are in the realm of the general theory of relativity, and not the special theory. The special theory of relativity assumes that space and time are rigid in terms of a background manifold. In the general theory, the underlying space-time manifold is dynamic and fluid; as such, space can curl up onto itself, and into various shapes, producing the bubble voids, which can expand a-causally.

Inflation also requires introducing a hypothetical particle, the inflaton, which drives the process [8-10]. Inflation is necessary, among other reasons, to explain the homogeneity, and at the same time the perturbations, associated with WMAP/Planck CBR temperature profile observed at photon-matter decoupling, which happened at roughly 380,000 years after the big bang. Inflation with the inflaton field, we will argue, may not be needed. Thermal quantum fluctuations caused by virtual particle creation and annihilation may ultimately be responsible for the perturbations in temperature, which freeze out.

In standard cosmology, one typically assumes adiabatic expansion of the universe as a whole with maximum entropy increase where we characterize the earliest times by seemingly infinite temperatures, pressures and a singular, i.e., zero volume. Perhaps this is not true. Perhaps there is a finite cut-off for all these variables, and the universe need not expand forever, as is the case for negative or zero curvature. A possibility is that the cosmos will eventually contract and experience a big bounce. The energy, which is needed for cosmic expansion, may not be entirely adiabatic  The cosmos may have developed another way; perhaps part of the evolution of the universe was isothermal. Part of that expansion might have been heat input from one part of the universe to another within a given time period, the inflation phase.

We will model the expansion of the universe as a thermodynamic heat engine, specifically, as a Carnot process (cycle) where we have isothermal expansion from points a → b, adiabatic expansion from b → c, isothermal contraction from c → d, and adiabatic contraction from d → a. At point a, we have a very high but *finite* pressure and temperature, and a non-zero volume. At point b, adiabatic expansion begins, and this is the phase where the cosmos presently finds itself. At point c, the universe will begin isothermal contraction having exhausted all of its heat energy, as well as its internal energy, required for further expansion. And at point d, the final phase begins where the universe further contracts, but adiabatically to its original starting point, point a, from which a big bounce may occur.

As in any heat engine, we extract heat from a hotter reservoir and expel into a colder one. The universe being a "closed system" in a thermodynamic sense has to contain both its "system" and its "surroundings" within its confines. We define our system collectively as those regions of space, within the observable universe, which will eventually evolve into voids and empty space. It is only in these regions that cosmic expansion takes place. These pockets evolved from a slightly elevated temperature in the WMAP CBR of the order $\Delta T/T \sim 5 * 10^{-5}$ where T is the temperature. We will define the surroundings, collectively, as those regions within the observable universe, which will eventually evolve into massive galaxies, clusters, super-clusters and filaments. These are regions, i.e., pockets characterized by little to no expansion due to the action of gravity, and for which we have slightly lower temperatures in the WMAP profile, $\Delta T/T \sim -5 * 10^{-5}$. The idea is that the surroundings will fuel the expansion of the system in the isothermal phase a → b, through heat input (inflation phase) keeping the universe as a whole a "closed system". Then in the adiabatic phase, which follows, change in internal energy will force subsequent expansion. We chose a Carnot cycle as an ansatz; it is well-known that it has maximum efficiency, is a reversible process, and the ratio of heat expelled to heat extracted over a complete cycle is expressible solely in terms of temperatures of the reservoirs, i.e., $|Q_C|/|Q_H| = T_C/T_H$. Heat energy $|Q_H|$, is transferred into the system from the surroundings in transitioning from a → b, causing isothermal expansion. Heat energy, $|Q_C|$, is given up to the surroundings resulting in isothermal contraction, from points c → d.

We define both system and surroundings, at very high temperatures, in terms of blackbody radiation, which we believe is the primordial form of energy in cosmic evolution [11-14]. We drive the equations of state for the four processes listed above. We will show that the efficiency for a blackbody radiation cycle is calculated by the same formula as that for an ideal gas in a Carnot cycle, i.e., $e_{CARNOT} = (1 - T_C/T_H)$. This result has been confirmed independently in another and earlier work [15], in the context of defining thermodynamic variables for a blackbody in n-spatial dimensions. We also give the total work done which will drive the expansion/contraction process over a complete cycle, $W = |Q_H| - |Q_C|$, which also confirms a previous result [16]. The total work over a complete cycle will be expressed in terms of temperature, $T_C$ and $T_H$, and volume increase and decrease.

The organization of this paper is as follows. In part II, we introduce the thermodynamic variables of interest for a blackbody. The derivation follows the significant work of reference [17]. We define the Carnot cycle in terms of these variables, and give the efficiency and work done. We apply this scheme to the universe and work out some of the consequences. In section III, we present arguments for an alternative view of inflation. Isothermal expansion starting from a fixed pressure and volume will allow for a relatively homogeneous temperature distribution. Thermal quantum fluctuations will freeze out in magnitude at very high temperatures and allow for in-homogeneities in temperature, which some 380,000 years afterwards, will be spatially localized upon matter-photon decoupling. This will reproduce what we observe in WMAP. Freeze-out is caused by isothermal expansion, and heat transfer is a one-way street from surroundings to system because the system, upon expanding, has lost its energy for further heat transfer back to the surroundings. Thermal equilibrium is thereby lost for the universe as a whole but maintained within the boundaries of what we define as our "system". We calculate the temperature of freeze-out for $\Delta T/T$ and explain how entropy increase leads to an arrow of time in the inflation phase. In part IV, we present our summary and conclusions.

## II. Thermodynamics of the Early Universe

In 3-dimensional space, we know [18-22] that the number of nodes per unit frequency for a blackbody is given by $(8\pi V)(\nu^2/c^3)$, where $\nu$ is the frequency and V is the volume. We multiply the above by the probability of finding a photon with frequency between $\nu$ and $(\nu + d\nu)$, the Bose-Einstein factor. This gives the number of photons, dN, in that particular frequency range as

$$dN = (8\pi V)(\nu^2 d\nu / [c^3 (e^{h\nu/kT}-1)]) \qquad (1)$$

Integrating $(h\nu) dN/V$ over all frequencies gives the energy density per unit volume, $u = u(T)$. The result is

$$u(T) = (8\pi^5/15)(kT)^4/c^3h^3$$

$$= 4\sigma T^4/c \tag{2}$$

The Stefan-Boltzmann constant $\sigma$ has the numerical value, $5.67 * 10^{-8}$ Watts/ ($m^2 * K^4$).

The internal energy U, entropy S, pressure p, and Helmholtz free energy F, defined as $F \equiv (U - TS)$, can all be expressed in terms of $u = u(T)$. One finds [23,24]

$$U = uV, \quad S = (4/3)(uV/T), \quad p = u/3, \quad F = -uV/3 \tag{3a,b,c,d}$$

The entropy density, s, and Helmholtz free energy density, f, are found by factoring out the volume, V, in equations (3b) and (3d).

We first consider isothermal expansion. Since temperature, T, is a constant, both the energy density, u, and pressure, p, are constant. Hence $U = uV$ will be proportional to volume V, and

$$U_2/U_1 = (a_2/a_1)^3 \quad \text{(isothermal expansion)} \tag{4}$$

In equation (4), "a" is the cosmic scale parameter in Hubble's law ($\dot{a} = H a$), and the subscripts 1, 2 refer to different cosmic times, i.e., epochs. Similar relations hold for S and F since these quantities are also proportional to volume. It is clear that

$$S_2/S_1 = (a_2/a_1)^3, \quad F_2/F_1 = (a_2/a_1)^3 \quad \text{(isothermal expansion)} \tag{5a,b}$$

The first law of thermodynamics can be written as $dQ = dW + dU$ where $dQ$ is the heat entering the system, $dW$ the work done by the system, and $dU$ is the change in the internal energy of the system. Because the temperature is a constant, $dU = d(uV) = udV$. Also, $dW = pdV = uV/3$ by equation (3c). Therefore,

$$dQ = TdS = pdV + udV$$

$$= 4u \, dV/3$$

$$= 4p \, dV \quad \text{(isothermal expansion)} \tag{6}$$

Integrating equation (6), we obtain

$$Q_2 - Q_1 = 4/3 \, u \, (V_2 - V_1)$$

$$= 4 \, p \, (V_2 - V_1) \quad \text{(isothermal expansion)} \tag{7}$$

Thus, in transitioning from $a \rightarrow b$ in our Carnot cycle,

$$Q = |Q_H| = 4/3 \, u_H \, (V_b - V_a) \quad \text{(isothermal expansion)} \tag{8a}$$

And upon isothermal volume contraction from $c \rightarrow d$,

$$Q = -|Q_C| = 4/3 \, u_C \, (V_d - V_c) \quad \text{(isothermal contraction)} \quad (8b)$$

The quantities, $u_H$ and $u_C$, are the energy volume densities defined by equation (2) where we use $T_H$ for the temperature of the hotter reservoir, and $T_C$ for the temperature of the colder reservoir. The total work done by this heat engine in completing one cycle is therefore

$$W_{TOTAL} = |Q_H| - |Q_C|$$

$$= 4/3 \, u_H \, (V_b - V_a) - 4/3 \, u_C \, (V_c - V_d) \quad (9)$$

$$= 4 \, p_H \, (V_b - V_a) - 4 \, p_C \, (V_c - V_d) \quad \text{(isothermal expansion/contraction)}$$

The change in internal energy is zero since we are back at our starting point. Notice that for a change in volume, the pressure is a constant; thus, an *isothermal* process is also, at the same time, an *isobaric* one for thermal radiation.

Adiabatic expansion and contraction are next considered. Since dQ equals zero, we find from the 1st law of thermodynamics, that

$$0 = p \, dV + dU \quad \text{(adiabatic process)} \quad (10)$$

We integrate this to find the total work done in expansion. Unlike isothermal expansion where the volume expansion is driven by heat, in adiabatic expansion it is caused by a decrease in internal energy. We re-write equation (10) as follows:

$$p \, dV = - d(uV) = -3 \, (dp \, V + p \, dV) \quad \text{(adiabatic process)}$$

, where we have made use of $p = u/3$ for the second equality. From this, it follows from equation (10) that

$$4p \, dV = -3 \, dp \, V \quad \text{(adiabatic process)}$$

This is straightforward to integrate; the result is $p_2/p_1 = (V_1/V_2)^{4/3}$ or, alternatively,

$$p_1 \, V_1^{4/3} = p_2 \, V_2^{4/3} \quad \text{(adiabatic process)} \quad (11)$$

Our adiabatic equation of state gives $\gamma = 4/3$, which is to be expected for blackbody radiation. Equation (11) is be contrasted with equation (7), which holds for an isothermal process.

In an adiabatic process, we can furthermore claim that $T_1^4 \, V_1^{4/3} = T_2^4 \, V_2^{4/3}$. This follows from equation (11) upon realizing that p is proportional to u, and u, in turn, is proportional to $T^4$. See equation (2). Taking a quartic root, we obtain

$$T_1 \, V_1^{1/3} = T_2 \, V_2^{1/3} \quad \text{(adiabatic process)} \quad (12)$$

Equation (12) is important because it allows us to establish a fundamental relation in cosmology, namely, that temperature increase is inversely proportional to scale parameter decrease. We know that

$$T/T_0 = (V_0/V)^{1/3} = a_0/a = (1+Z) \quad \text{(adiabatic process)} \quad (13)$$

However, *this relation holds only for an adiabatic process, and not for an isothermal one*! Volume expansion, as well as the evolution of time, in the isothermal phase will be very different, as we shall see. The subscript "0" stands for the present cosmological epoch, "a" is the cosmological scale parameter and Z is the redshift. At temperature T, the scale parameter is "a". The exponent $\gamma = 4/3$ in equation (11) is critical in establishing the relationship given by equation (13).

Finally, since p is proportional to u, we see that

$$p_1 T_1^{-4} = p_2 T_2^{-4} \quad \text{(adiabatic process)} \quad (14)$$

A decrease in temperature leads to the familiar dramatic fourth power decrease in radiative pressure. In contrast, *for an isothermal process, both temperature and pressure remain constant*.

Coming back to our Carnot cycle, in going from point b to point c, we expect

$$p_b V_b^{4/3} = p_c V_c^{4/3} \quad \text{(adiabatic process)} \quad (15)$$

And also,

$$p_b/p_0 = (a_0/a_b)^4 = (1+Z_b)^4 \quad \text{(adiabatic process)} \quad (16)$$

In transitioning from points d → a, we must have

$$p_d V_d^{4/3} = p_a V_a^{4/3} \quad \text{(adiabatic process)} \quad (17)$$

Here, the volume $V_a$ is less than $V_d$, and, consequently, $p_a$ is greater than $p_d$. It can also be established from equation (12) that

$$T_b V_b^{1/3} = T_c V_c^{1/3}, \quad T_d V_d^{1/3} = T_a V_a^{1/3} \quad \text{(adiabatic process)} \quad (18a,b)$$

Since V is proportional to the cosmic scale parameter cubed, this allows us to write

$$T_b/T_c = T_H/T_C = a_c/a_b, \quad T_a/T_d = T_H/T_C = a_d/a_a \quad (19a,b)$$

We employ the relations above to construct our Carnot cycle. In terms of a qualitative p-V diagram, we obtain figure 1 shown below:

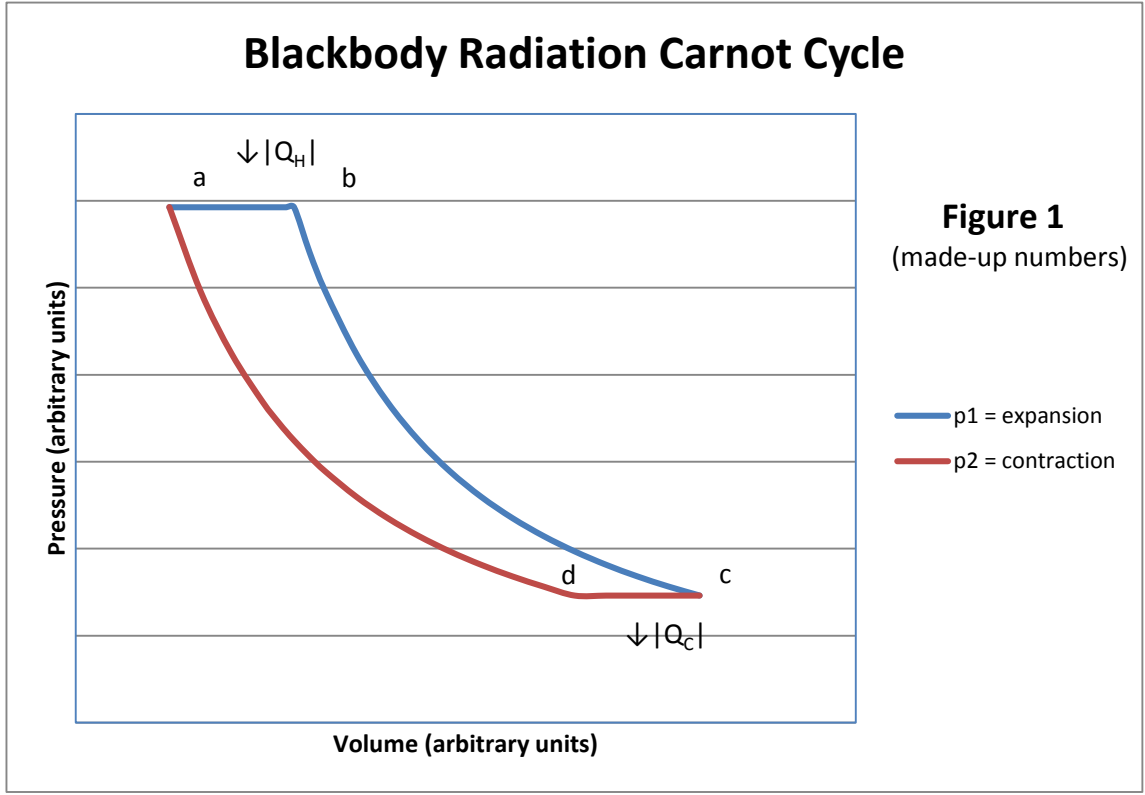

In figure 1, the lines drawn from points a to b, and from points c to d, are drawn greatly exaggerated, lengthwise, in this diagram. They should be drawn almost infinitely close to one other if this figure were to scale, which it is not in either the x or y sense. Figure 1 is for qualitative illustrative purposes only. Even though the numbers are made up for this illustration, the relations (15) and (17) are satisfied.

We remark that $p_H = u_H/3 = 4\sigma T_H^4/(3c) =$ constant with $T_a = T_b = T_H$ for the phase a → b. Similar relations hold for the transition c → d; $p_C = u_C/3 = 4\sigma(T_C)^4/(3c) =$ constant where $T_c = T_d = T_C$. Heat energy, in the amount $|Q_H|$, is extracted from the surroundings from a → b; it will be released to the surroundings in c → d, in the amount $|Q_C|$ much, much later. The total work done in expanding and contracting 3-dimensional space is the area enclosed in figure 1, and it is specified mathematically by equation (9). More explicitly, $|Q_H|$ is what fuels the expansion from a → b. A decrease in internal energy forces further expansion from b → c. At point c, all heat and internal energy will have been exhausted and contraction begins. Heat loss drives contraction from c → d, and an increase in thermal internal energy drives further contraction from thermodynamic points d → a, thus completing the cycle. We will identify the isothermal process from a → b as the inflation phase, shortly, in section III. The isothermal contraction process from c → d would be recognized as the deflation phase.

We consider the efficiency of our heat engine next. The efficiency is defined conventionally as

$$e_{CARNOT} \equiv W/|Q_H| = 1 - |Q_C|/|Q_H|$$

$$= 1 - (T_C/T_H)^4 ((V_c - V_d)/(V_b - V_a)) \quad (20)$$

For the second line, equations (8) have been utilized. But for adiabatic expansion and contraction, we know from equations (18) that

$$T_H V_b^{1/3} = T_C V_c^{1/3}, \qquad T_C V_d^{1/3} = T_H V_a^{1/3}, \quad (21a,b)$$

This allows us to further claim that

$$((V_c - V_d)/(V_b - V_a)) = (T_H/T_C)^3 \quad (22)$$

Thus, by substituting equation (22) into equation (20), we have our result,

$$e_{CARNOT} = (1 - T_C/T_H) \quad (23)$$

This is a familiar result because it also holds for an ideal gas; we see that for blackbody radiation the efficiency is calculated in an analogous manner. Because $T_H \ggg T_C$, we expect close to 100% efficiency because of equation (23). Ultimately, the efficiency rests on the deviation from flatness, i.e., on the curvature of the universe. This is so because the amount of positive curvature will determine at the very end how close to zero our final temperature, $T_C$, gets. For zero or negative curvature, i.e., for an open universe, $T_C$ will approach zero, but then we would not have a heat engine, nor would we have an efficiency.

To find the total work done over a cycle, we simplify the expression, equation (9), by using equation (22). For equation (9), we now find

$$W_{TOTAL} = |Q_H| - |Q_C|$$

$$= 4 p_H (V_b - V_a) - 4 p_C (V_c - V_d)$$

$$= 4 p_H (V_b - V_a) - 4 p_C (T_H/T_C)^3 (V_b - V_a)$$

$$= 4 p_H (V_b - V_a) [1 - (p_C/p_H)(T_H/T_C)^3]$$

$$= 4 p_H (V_b - V_a) (e_{CARNOT}) \quad (24)$$

Remember that p is proportional to the temperature to the fourth power, and so, we are able to derive the fourth line in equation (24) from the third. Since the efficiency is almost unity, we need really only know $V_a$, $V_b$, and $T_H$, the temperature of the hot reservoir for an evaluation of $W_{TOTAL}$. We will give an estimate for $|Q_H|$, and $T_H$, in the next section.

### III. An Alternative View of Inflation

In this section, we present arguments for why the isothermal process a → b may be a more natural ersatz for inflation as it does not require a hypothetical field, the inflaton. In addition, causality is clearly defined as applying to material bodies, the surroundings, and not the system, which are the voids. It is only the voids, which will expand. WMAP can be reproduced in its essentials with respect to the homogeneity in temperature, and the inhomogeneity in $\Delta T/T$. The angular size of separation between the hot and cold regions, about one degree on average in angular resolution as shown in the power spectrum, will be considered in another paper. The results of this section actually stand independently in the sense that we do not have to assume a heat engine, nor a universe with necessarily re-collapses. In other words, the results will hold irrespective of whether we are dealing with a closed, flat or open universe.

We first estimate the freeze-out temperature for $\Delta T/T$ using the WMAP observed result, $\Delta T/T \sim 5 * 10^{-5}$. We start with the indeterminacy principle of Heisenberg between energy, E, and time, t. Since $\Delta E = k\Delta T$, we can write $k\Delta T \, \Delta t \geq h/4\pi$, where h and k are Planck's and Boltzmann's constants, respectively. $\Delta T$ is the indeterminacy in temperature, and $\Delta t$ is the uncertainty in time. Since $\Delta t$ must be less than t, the age of the universe, it follows that

$$\Delta T/T \geq h/(4\pi kT) \, (1/t) \tag{25}$$

Therefore, for the thermodynamic points "a" and "b", we must have

$$(\Delta T/T)_a \geq h/(4\pi kT_a)(1/t_a), \qquad (\Delta T/T)_b \geq h/(4\pi kT_b)(1/t_b) \tag{26a,b}$$

We believe the indeterminacy in temperature is due to thermal quantum mechanical fluctuations, i.e., thermally driven radiative corrections (creation/annihilation of virtual particles). While a specific temperature determines a perfectly smooth blackbody spectrum by equations (1) and (2), fluctuations in T, $\Delta T$, produce fluctuations in photon frequency, $k\Delta T = h\Delta\nu$, where $\nu$ is a typical frequency in the blackbody distribution, given by equation (1).

We next wish to relate expansion time to temperature. We first consider adiabatic expansion which holds within the range $t_c \geq t \geq t_b$. In general, $H = (\dot{a}/a)$, where "a" is the cosmic scale parameter and H is Hubble's constant. Therefore,

$$dt = da/(H\,a) \tag{27}$$

However, because of the Friedmann equation, H can be written as $H = (8\pi G \, u/3c^2)^{1/2}$. We are in a thoroughly radiation-dominated phase where the energy density, $\rho c^2$, is given exclusively by black body radiation, u. We substitute this expression for H into equation (27) to obtain

$$dt = (B/u)^{1/2} \, da/a \tag{28}$$

In equation (28), the constant B has been defined as $B \equiv (3c^2/8\pi G) = 1.61 * 10^{26}$ kg/m. Next, according to equation (13), $T/T_0$ equals $a_0/a = (1+Z)$ for an adiabatic process where the subscript "0" refers to the present epoch, T and "a" are arbitrary CBR temperatures and associated scale parameter, respectively, and Z is the corresponding redshift. From equation (13), it follows directly that $dt/t = - da/a$, and thus

$$dt = - (B/u)^{1/2} \, dT/T \tag{29}$$

This we integrate, where u is given by equation (2), and obtain

$$\int dt = - \int A T^{-3} \, dT$$

$$(t - t_b) = \tfrac{1}{2} A T^{-2} - \tfrac{1}{2} A T_b^{-2} \qquad (t_c \geq t \geq t_b) \tag{30}$$

In equation (30) we have defined another constant A, where $A \equiv (Bc/4\sigma)^{1/2} = 4.62 * 10^{20}$ $K^2$ s. From equation (30), it can be deduced that

$$t_b = \tfrac{1}{2} A T_b^{-2} \tag{31}$$

$$= 2.31 * 10^{20} \, (K^2 \, s) \, T_b^{-2}$$

One way to justify this relation is to assume the converse, $t = t_b \neq \tfrac{1}{2} A T_b^{-3}$, and notice that then, equation (30) would lead to a contradiction.

Equation (31) is the equation we will use to eliminate time, $t_b$, in equation (26b). Substituting (31) into (26b) renders, after a cancellation in $T_b$,

$$(\Delta T/T)_b \geq h/(2\pi A k) \, T_b \tag{32}$$

This equation indicates that the greater a given threshold background temperature, the greater the relative fluctuations in temperature we should expect. Furthermore, for $(\Delta T/T)_b$, we may substitute our value of $\Delta T/T \sim 5 * 10^{-5}$. WMAP is a spatial distribution of temperature, which was formed very early in the universe at incredibly high temperatures. This we identify with point b. As radiation decoupled from matter many years later, some 380,000 years later, the temperature profile spatially "froze out" and is the one we observe today. Substituting our $(\Delta T/T)_b \sim 5 * 10^{-5}$ value into relation (32) and evaluating all constants provides us with the following estimate for $T_b$:

$$T_b \sim 3.02 * 10^{27} \, K \tag{33a}$$

This, we believe, is the temperature of freeze-out for $\Delta T/T \sim 5 * 10^{-5}$. It happened at a time when the universe was in the earliest time of formation; by equation (13) we estimate the associated time to be

$$t_b \sim 2.54 * 10^{-35} \text{ s} \tag{33b}$$

This marks the end of the isothermal expansion phase, which we identify as the end of inflationary epoch. The values indicated by equations (33a,b) are in general agreement with current estimates for the end of the inflationary period (between $10^{-36}$ and $10^{-32}$ s with corresponding temperatures) in standard cosmology.

We note that equation (33a), in particular, enables us to estimate the energy density and pressure at thermodynamic point b. Employing equations (2), (3c), and substituting the temperature from equation (33a), we obtain

$$u_b \sim 6.29 * 10^{94} \text{ J/m}^3, \qquad p_b \sim 2.10 * 10^{94} \text{ N/m}^2 \tag{34a,b}$$

These estimates for background energy density and pressure not only hold at point b; they also hold at point a, and every point in between as we are dealing with an *isothermal* process in going from a $\rightarrow$ b. Remember that $T_a = T_b$.

We next consider isothermal expansion in more detail from point "a" to point "b", a $\rightarrow$ b. First, we recognize that we cannot use equation (30) since time would literally stand still for a temperature, which is constant. Therefore, we *must derive a new relation for expansion time, independent of temperature and different from equation (30)*. The equation (28) is our starting point. What drives volume expansion in this isothermal phase is not temperature, or its' proxy which is internal energy. Rather it is heat taken in by the system from the surroundings, which causes volume increase. Therefore, we replace the internal energy density, u, in equation (28) by heat energy density, q, defined by $q \equiv Q/V$, as this forces expansion from a $\rightarrow$ b. Equation (6) tells us that $dQ = T\, dS = 4p\, dV$. In addition, because temperature is a constant and p depends only on temperature, we can integrate this first law to obtain $Q = 4p\, V$. Thus, our ersatz for equation (28) becomes

$$dt = (BV/Q)^{1/2}\, da/a$$

$$= (B/4p)^{1/2}\, da/a \tag{35}$$

, where $p = u/3 = u_b/3 = $ constant. See equations (34a) and (34b).

We integrate equation (35) between points a and b and obtain the result,

$$(t_b - t_a) = (B/4p)^{1/2} \ln(a_b/a_a)$$

$$= 1/3\, (B/4p)^{1/2} \ln(V_b/V_a) \tag{36}$$

Remember that volume V is proportional to $a^3$ where "a" is the cosmic scale parameter. Finally, recognizing that $A = (Bc/4\sigma)^{1/2}$ and $p = u/3$, we find for equation (36),

$$(t_b - t_a) = 1/3\ (3\ B/4u_b)^{1/2}\ \ln(V_b/V_a)$$

$$= (2\ \sqrt{3})^{-1}\ AT_b^{-2}\ \ln(V_b/V_a)$$

$$= (\sqrt{3})^{-1}\ t_b\ \ln(V_b/V_a) \tag{37}$$

In the last line, we have made use of equation (31). We compare equation (37), which holds for an isothermal process, with equation (30), which holds for an adiabatic process. Both are clearly very different in predicting how time evolves. We notice that equation (37), in particular, does not involve temperature.

We next estimate the ratio, $(V_b/V_a)$. We divide equation (27b) by (27a) to obtain

$$(\Delta T/T)_b\ /\ (\Delta T/T)_a\ =\ t_a/\ t_b \tag{38}$$

Remember that $T_a = T_b$ but $V_a \neq V_b$. The maximum fluctuation for a given temperature is $(\Delta T/T) \sim 1$; this we identify as point a. Thus, $(\Delta T/T)_a \sim 1$. Inserting this, and our original premise, $(\Delta T/T)_b \sim 5 * 10^{-5}$, into equation (38) gives

$$t_a/\ t_b = 5 * 10^{-5} \tag{39}$$

However, from equation (37), we see

$$(1 - t_a/\ t_b\ ) = (\sqrt{3})^{-1}\ \ln(V_b/V_a) \tag{40}$$

Upon substitution of equation (39) in equation (40), we can solve for this ratio of volumes, and find

$$V_b/\ V_a = 5.65 \tag{41}$$

This is a very small increase in volume, but keep in mind that the period is exceedingly small. This increase in volume holds for the voids, which is that portion of the universe, which undergoes expansion. Equation (41) substituted into equation (37) will give us the time of inflation.

To find that time period, $(t_b - t_a)$, we can use equation (37) with equations (33b) and (41) substituted; the result is

$$(t_b - t_a) = 2.539873 * 10^{-35}\ s \tag{42}$$

This is slightly less than $t_b = 2.54 * 10^{-35}$ s, equation (33b). An alternative is to make use of equation (39) with (33b). This gives $t_a = 1.27 * 10^{-39}$ s, which, incidentally, is significantly larger than the Planck time of $5.39 * 10^{-44}$ s. We can subtract $t_a$ from (33b), also giving equation (42). Either way, this brings us back to our initial starting point, point a, where $t_a = t_b - (t_b - t_a) = 1.27 * 10^{-39}$ s. See relation (33b). We emphasize that $t_a$ is the beginning of volume expansion, and not the beginning of cosmological time. For a cycle, there is no beginning or end to time.

Quantum fluctuations in temperature are an integral part of time evolution in the isothermal phase, a → b. If we relied on equation (30), time would stand still in this phase, as there is no difference in background temperature. We made use of equation (36) because even though temperature is constant, the change in volume allows for an evolution in time. The underlying reason for this change in volume is entropy increase and quantum fluctuations. Quantum disturbances allow for heat flow from surroundings to system causing volume expansion of the latter and producing slightly depressed temperatures in the former. Infinitesimally, from surroundings to system, the first line of equation (6) gives the heat being absorbed, where the quantity, dS, represents the change in entropy. The heat flow, with the accompanying *increase* in entropy, create a one-way street for energy flow because once the system utilizes its incoming heat for expanding by an amount dV, it can no longer transfer heat back to the surroundings. The increase in entropy defines an arrow of time, which we cannot reverse.

In the analysis above, we assume that the gravitational constant is a true constant of nature, i.e., it does not change in value all the way back to point a. We note that we defined our constants, A and B, specifically in terms of G. Moreover, if one accepts the heat engine model, there is no beginning of cosmological time, due to the cyclical nature of expansion/contraction, as was already mentioned.

Finally, let us estimate the heat input in the isothermal era. Because $p_H = u_b/3$, and because of equation (41), equation (24) can be rewritten as

$$|Q_H| = 4/3 \; u_b \; (V_b - V_b/5.65)$$

$$= 1.1 \; U_b \qquad (43)$$

In this equation, $U_b = u_b V_b$ is the internal energy at point b. We can also claim that

$$U_b/U_a = (V_b/V_a) = 5.65, \quad S_b/S_a = (V_b/V_a) = 5.65, \quad F_b/F_a = (V_b/V_a) = 5.65 \qquad (44a,b,c)$$

These relations hold because of equations (4) and (5). Internal energy, entropy, and Helmholtz free energy have increased by a factor 5.65 when transitioning from points a → b.

We will now estimate the heat input in the isothermal (inflation) phase. We know that relation (13) can relate the volume at point b to the current volume of the observed universe. Because we are still in the adiabatic phase,

$$V_b = (T_0/T_b)^3 \; V_0 = 7.35 * 10^{-82} \; V_0 \qquad (45)$$

As always, the subscript "0" stands for the present epoch. We know that currently, $T_0 = 2.725$ K, and relation (33a) specifies $T_b$. We substitute equations (45) and (34a) into relation (43) to obtain

$$|Q_H| = 5.08 * 10^{13} \ V_0 \qquad (46)$$

In equation (46) the heat input, $Q_H$, is measured in Joules and the volume, $V_0$, in m$^3$.

The present size of the observable universe is estimated [25,26] to have a diameter equal to $8.8 * 10^{26}$ m. Using this estimate, we ascertain a present epoch volume of roughly $V_0 = 4\pi/3 \ R^3 = 3.56 * 10^{80}$ m$^3$. This we substitute in both (45) and (46), respectively, and find

$$V_b = .262 \text{ m}^3, \qquad |Q_H| = 1.81 * 10^{94} \text{ J} \qquad (47a,b)$$

Equation (47a) gives an equivalent aggregate diameter of about .80 m for the voids. By equation (41), $V_a = .046$ m$^3$ and the matching equivalent, collective diameter is .45 m. This volume increase is significantly less than the trillion, trillion-fold expansion, in less than a trillionth of a trillionth of a second.

We close this section with one final remark. If we knew the temperature at point c, we could use the results of the previous section to determine $u_c$, $p_c$, $V_c$, etc. at point c. Equation (11), in particular, tells us that

$$u_b \ V_b^{4/3} = u_c \ V_c^{4/3} \qquad (48)$$

We have estimates for $u_b$ and $V_b$ and if we knew $T_c$, we could determine $u_c$. That would enable us to find the maximum extent of volume expansion, $V_c$, by using equation (48). The converse also holds true; if we could estimate the volume, $V_c$, we could determine $T_c$. Unfortunately, we know neither. Furthermore, if the universe were open or flat, then $V_c$ would have no limit. Relation (48) would still hold, but then, as $V_c^{1/3} = a_c$ approaches infinity, the temperature, $T_c$, would approach zero. Space would expand forever without limit, and the blackbody temperature would tend to zero.

IV. **Summary and Conclusion**

We modeled the expansion of the universe as a heat engine, where we extract heat from one part of the universe, the "surroundings", and give it up to another part of the universe, the "system". This heat fuels the initial volume increase of the cosmos, which we identify as the inflationary phase. The universe is thought to follow a Carnot cycle where we have isothermal expansion from a → b (inflationary era), adiabatic expansion from b → c, isothermal contraction from c → d, and, finally, adiabatic contraction from d → a. See figure (1). In a → b, the energy density and pressure stay relatively constant, but thermal fluctuations cause heat flow which are utilized by the "system" to increase volume. Upon volume expansion, the system has no heat energy left over to transfer back to the surroundings, and thus a one-way street is established where entropy can only increase. We have a 5.65-fold increase in entropy in transitioning from point "a" to point "b" by equation (44b). Internal energy also increases by a factor of 5.65, as seen by equation (44a). From b → c, further volume expansion is produced, but this time due to a different mechanism, a decrease in internal thermal energy. This adiabatic expansion phase is where the universe currently finds itself. After all the internal energy is used up, we postulate a contraction, first isothermally from c → d (deflation phase), and then adiabatically from d → a. From c → d, the universe contracts due to the system giving off heat energy, $|Q_C|$, to the surroundings. In addition, from d → a, internal thermal energy is increased to effect a final volume contraction closing the cycle.

We have assumed a universe which closes, i.e., a universe with very slight positive curvature such that we can allow for a big bounce. The results of section III, however, allow us to relax this requirement. In the inflationary phase, all results, which we derive, will remain valid irrespective of the signature of the curvature of space. We defined the system collectively as those regions of space within the observable universe, which have slightly elevated temperature in the WMAP profile. These are the regions or pockets, which will eventually expand, and later evolve into cosmic voids and empty space. We define the surroundings collectively as those regions of space within the observable universe, which have slightly lower temperatures in the WMAP profile. These are the patches or pockets, which will not expand. The build-up of matter will occur within those regions, which have slightly depressed temperatures. These regions will later evolve into galaxies, clusters, super-clusters and filament walls. We assume that the magnitude of the temperature fluctuations in WMAP occurs very early in cosmological time, within the first second. What we see in WMAP are the remnants, which froze out spatially after photon-matter decoupling, roughly 380,000 years after first formation. If gravity remained constant, the $\Delta T/T$ magnitude freeze-out temperature, and freeze-out time, will both be specified by equations (33a) and (33b), respectively.

The advantages of this heat engine model are many; within this framework, we can

1) Provide a specific process, i.e., mechanism for what drives volume expansion, and contraction, over a complete cycle.
2) Avoid a singularity in volume, and prevent infinite temperatures, pressures and energies from arising at the beginning of expansion. The universe may have had a finite size at the beginning of evolution.
3) Develop expressions for the total work done, the efficiency, internal energy, pressure, and entropy. See, for example, equations (34a) and (34b). The energy density and pressure stay relatively constant in the inflationary period, from a → b. They will also stay relatively constant in the deflationary phase, from c → d, if we believe in a closed, i.e., positively curved universe. That heat input during the inflation phase is estimated to be $|Q_H| = 1.81 * 10^{94}$ Joules.
4) Estimate the temperature, pressure, volume and time for $\Delta T/T$ formation, at point b. See equations (33a), (33b), (34b) and (47a). Numerically, $T_b = 3.02 * 10^{27}$ K, $p_b = 2.10 * 10^{94}$ N/m$^2$, $V_b = .262$ m$^3$, and $t_b = 2.54 * 10^{-35}$ s.

5) Estimate the temperature, pressure, volume and time at the start of the inflationary period, point a. See the discussion following equation (34b), and equations (39) and (41). $T_a = T_b$ but $V_a = V_b / 5.65$. Numerically, $T_a = 3.02 * 10^{27}$ K, $p_a = 2.10 * 10^{94}$ N/m$^2$, $V_a = .046$ m$^3$, and $t_a = 1.27 * 10^{-39}$ s.
6) Entertain an alternative framework for inflation, one which has a fixed constant background temperature, has a clear definition for causality (system expands a-causally; surroundings do not) and does not require a hypothetical field, the inflaton.
7) Give a cosmic time evolution, which holds for isothermal expansion, and which differs markedly from that for adiabatic expansion. See equations (37) and (40), which indicate a different type of time evolution during inflation.
8) Finally, this model allows for a possible big bounce scenario where the universe can reconfigure itself at point a, due to the high temperature cauldron.

If we take our heat engine model seriously, then the positive curvature of space has to be established. At present, we cannot be rule it out. An estimate for the curvature can help us estimate the final volume and temperature. See equation (48).

We close this summary with an observation. What could have triggered, i.e., initiated cosmic expansion in the first place at point a? That we do not know. Perhaps the underlying reason has to do with the dimensionality of space itself. As shown in reference [27], the dimensionality of space, n=3, was in all likelihood, no accident. It may have been decided upon, i.e., determined at a very early cosmological point in time by thermodynamics. Near n=3, the Helmholtz free energy density, f, has an inflection point, a maximum when plotted as a function of spatial dimension, n. This is the first of all the important thermodynamic variables (entropy density s, internal energy density u, pressure p, etc.) to reach a maximum value, and that maximum was reached right around n=3 with a temperature of approximately $T^* \approx .93 * T_{PLANCK} = 1.32 * 10^{32}$ K. Our estimate for $T_a$ is $T_a = T_b = 3.02 * 10^{27}$ K, equation (33a). These temperatures are not far off. While we may not know what happened between $T^* \approx 1.32 * 10^{32}$ K, and $T_a = T_b = 3.02 * 10^{27}$ K, we do know that space could only expand three dimensionally, once 3 dimensions was decided upon by nature. It should be obvious, therefore, that $T^* \geq T_a = T_b$, which it is.

The author wishes to thank the physics department at Gonzaga University for many helpful discussions and comments, in particular Professors Erik Aver, and Adam Fritsch. I would also like to thank Professor Robert Scherrer at Vanderbilt University for reviewing the manuscript. Final thanks goes to Mr. John Krehbiel, whose helpful questioning prompted me, in large part, to write this paper.

**References:**

[1] Tonry et al., (2003) Astrophys. J. 594 1-24

[2] Komatsu E., et al. [WMAP Collaboration], Astrophys. J. Suppl. **192** (2011) 18 [arXiv:0803.0547 [astro-ph]].

[3] C.L. Bennett; et. al., "Nine Year Wilkinson Microwave Anisotropy Probe (WMAP) Observations: Final Maps and Results". Astrophysical Journal Supplement 208(2): 20 (2013) arXiv:1212.5225

[4] Collaboration, Planck, PAR Ade, N Aghanim, C Armitage-Caplan, M Arnaud, et al., Planck 2015 results. XIII. Cosmological parameters. arXiv preprint 1502.1589v2[1] (https://arXiv.org/abs/1502.1589v2), 6 Feb 2015. The value quoted is $\Omega_k = -.005^{+.016}_{-.017}$ (95% Planck TT + low P + lensing). See section 6.2.4 on curvature, in particular, equation (49).

[5] A. Einstein (1917), Preuss. Akad. Wiss. Berlin Sitzber. 142-152

[6] J. A. Wheeler "Einstein's Vision" (Springer, Berlin, 1968).

[7] https://map.gsfc.nasa.gov/news (final 2 years of data) April 8, 2013

[8] A. H. Guth, "The Inflationary Universe; the Quest for a New Theory of Cosmic Origins" (1997)

[9] P.J. Steinhardt; N. Turok, "Endless Universe: Beyond the Big Bang", Random House, p114, ISBN 978-0-7679-1501-4 (2007)

[10] P.J. Steinhardt, "Inflation Debate; Is the Theory at the Heart of Modern Cosmology Flawed?" Sci. American, April 2011

[11] J. D. Barrow and W. S. Hawthorne. "Equilibrium Matter Fields in the Early Universe". Mon. Not. R. Astr. Soc, 243, (1990) 608-609.

[12] Barbara Ryden. "Introduction to Cosmology". Addison Wesley, Ohio EE. UU. (2006).

[13] [8] V Mukhanov, "Physical Foundations of Cosmology", (Cambridge University Press, USA, 2005).

[14] R. Brandenberger. "Introduction to Early Universe Cosmology". arXiv:1103.2271 [astro-ph.CO] (2011)

[15] Julian Gonzalez-Ayala, J. Perez-Oregon, Rubén Cordero and F. Angulo-Brown, "A possible cosmological application of some thermodynamic properties of the black body radiation in n−dimensional Euclidean spaces". Submitted.

[16] See reference [15], op. cit.

[17] Julian Gonzalez-Ayala, Rubén Cordero and F. Angulo-Brown, "Is the (3+1)-d Nature of the Universe a Thermodynamic Necessity?" EPL.DOI: 10.1209/0295-5075/113/40006. Also arXiv:1502.01843v2

[18] P. T. Landsberg and A. De Vos. "The Stefan Boltzmann constant in an n-dimensional space". Journal of Physics A: Mathematics and General, 22, (1989) 1073-1084. 4

[19] V. J. Menon and D. C. Agrawal. Comment on "The Stefan-Boltzmann constant in n-dimensional space". J. Phys. A: Math. Gen. 31, (1998) 1109-1110.

[20] See reference [11], op. cit.

[21] See reference [15], op. cit.

[22] See reference [12], op. cit.

[23] See reference [17], op. cit.

[24] See reference [15], op. cit.

[25] I. Bars and J. Terning, (November 2009). "Extra Dimensions in Space and Time". Springer. pp. 27–. ISBN 978-0-387-77637-8. Retrieved 1 May 2011.

[26] C. Lineweaver; T. M. Davis (2005), "Misconceptions about the Big Bang". Scientific American

[27] See reference [17], op. cit.